 \documentclass[final,5p,times,twocolumn]{elsarticle}

\usepackage{amssymb}
\usepackage{natbib}

\begin{document}

\begin{frontmatter}



\title{Miniaturized chip calorimeter for high-pressure cells at low temperature}


\author[first]{Neha Kondedan}
\author[first]{Andreas Rydh}
\ead{andreas.rydh@fysik.su.se}
\affiliation[first]{organization={Department of Physics},
            addressline={Stockholm University, Albanova University Center}, 
            city={SE-106 91 Stockholm},
            country={Sweden}}
            
\begin{abstract}
Heat capacity measurements under high pressure places high demands on the calorimeter. Here we describe the development of a miniaturized nanocalorimeter for high-pressure heat capacity measurements at low temperature. The device, fabricated on a silicon substrate, employs a high-frequency AC calorimetry technique and features a design with an outer diameter of 300~$\mu$m and thickness of 25-40~$\mu$m, small enough to fit into high pressure diamond anvil cells. Miniaturization is achieved by stacking all components, including thermometer and heaters, within a central area. The thin-film calorimeter thermometer measures 40~$\mu$m square and maintains the sensitivity and properties of larger thermometers. The fabrication process uses controlled anisotropic etch to produce calorimeter chips with a balance between robustness and thickness, suitable for experiments at high pressures and low temperatures. The calorimeter operates at a relatively high characteristic frequency between 10 Hz and 1 kHz, constraining the thermal oscillation to an effective volume dominated by the sample, thereby avoiding the use of a suspended membrane that is the basis for conventional nanocalorimeters.
\end{abstract}




\end{frontmatter}




\section{Introduction}
\label{introduction}
Specific heat is a crucial thermodynamic quantity for characterizing fundamental material properties and related phase transitions with respect to external conditions, such as magnetic field or pressure. Low-temperature specific heat measurements are particularly useful for understanding superconducting, magnetic, and structural transitions. The advent of AC calorimetry \cite{sullivan1968steady} and the development of various types of nanocalorimeters \cite{garden2012nanocalorimetry, yi2019nanocalorimetry, garden2009thermodynamics} have enabled specific heat measurements on samples down to the nanogram range. Thin-film nanocalorimeters have substantially reduced background heat capacity, but also opened for new measurement possibilities \cite{khansili2023calorimetric, willa2017nanocalorimeter, yi2017SEM, yi2015massspectrometry, zhang2005TEM}. 

There are various types of miniaturized calorimeters based on suspended mm$^2$ membranes combined with advanced microfabrication techniques. For substrate, some calorimeters are fabricated on 2-10~$\mu$m thick silicon substrates, taking advantage of the low heat capacity of silicon at low temperatures \cite{fominaya1997nanocalorimeter}. Others are constructed on an insulating polymer (poly-para-xylylene) with a copper frame \cite{bourgeois2007new, Lopeandia2010}. Most calorimeters are based on SiN$_x$ membranes \cite{denlinger1994thin, minakov2005thin, cooke2008thermodynamic, tagliati2012differential}. The critical component for the miniaturization is the thermometer. Some devices rely on thin-film thermocouples (mainly doped Si) \cite{minakov2005thin}, while most are resistive thin-film thermometers based on Pt \cite{cooke2008thermodynamic, denlinger1994thin}, NbN \cite{Lopeandia2010, bourgeois2005attojoule}, NbSi \cite{cooke2008thermodynamic, denlinger1994thin}, AuGe \cite{fon2005nanoscale, tagliati2012differential}, or, recently, NiCrSiO$_x$ \cite{fortune2023wide}.

Introducing high pressure to heat capacity measurements, especially at low temperatures, is of great interest because pressure directly affects the electronic density of states through the volume change, and can stabilize new phases and electronic states \cite{hermann2017chemical}.
However, high-pressure calorimetry remains a relatively unexplored area due to the significant experimental challenges associated with the integration of calorimeter and pressure cell. Diamond anvil cells are particularly desirable due to their ability to reach extremely high pressures \cite{jayaraman1983diamond, jayaraman1986ultrahigh}. But, while high-pressure cells in general, and diamond anvil cells in particular, require a small sample, such samples would need even smaller thermometers and heaters to not dominate the total measured heat capacity, thus excluding conventional adiabatic calorimeters.
For particular measurement ranges and samples, miniaturized calorimeters based on bare-chip thermometers or thermocouples, either inside or outside the pressure cell, have been devised \cite{jura1969technique, loriers1973specific, baloga1977ac, eichler1979method, bohn1991specific, demuer2000calorimetric, bouquet2000calorimetric, wilhelm2003ac, lortz2005evolution, kubota2008construction, umeo2016alternating, geballe2017ac, gati2019use, martelli2020high}. A high-pressure calorimeter for general use working over a broad temperature range and in magnetic fields is, to our knowledge, still missing.

Here, we report the fabrication and characterization of a ultra-miniaturized thin-film calorimeter on a silicon chip, that can be easily adapted to various high-pressure cells and cryostats. A key aspect of the fabrication process is the miniaturization of the thermometer while preserving its functional properties.

\section{Calorimeter fabrication}
\label{CALORIMETER DESIGN}

\subsection{Calorimeter design}

\begin{figure}[h!]
  \centering
\includegraphics[width=1.02\linewidth]{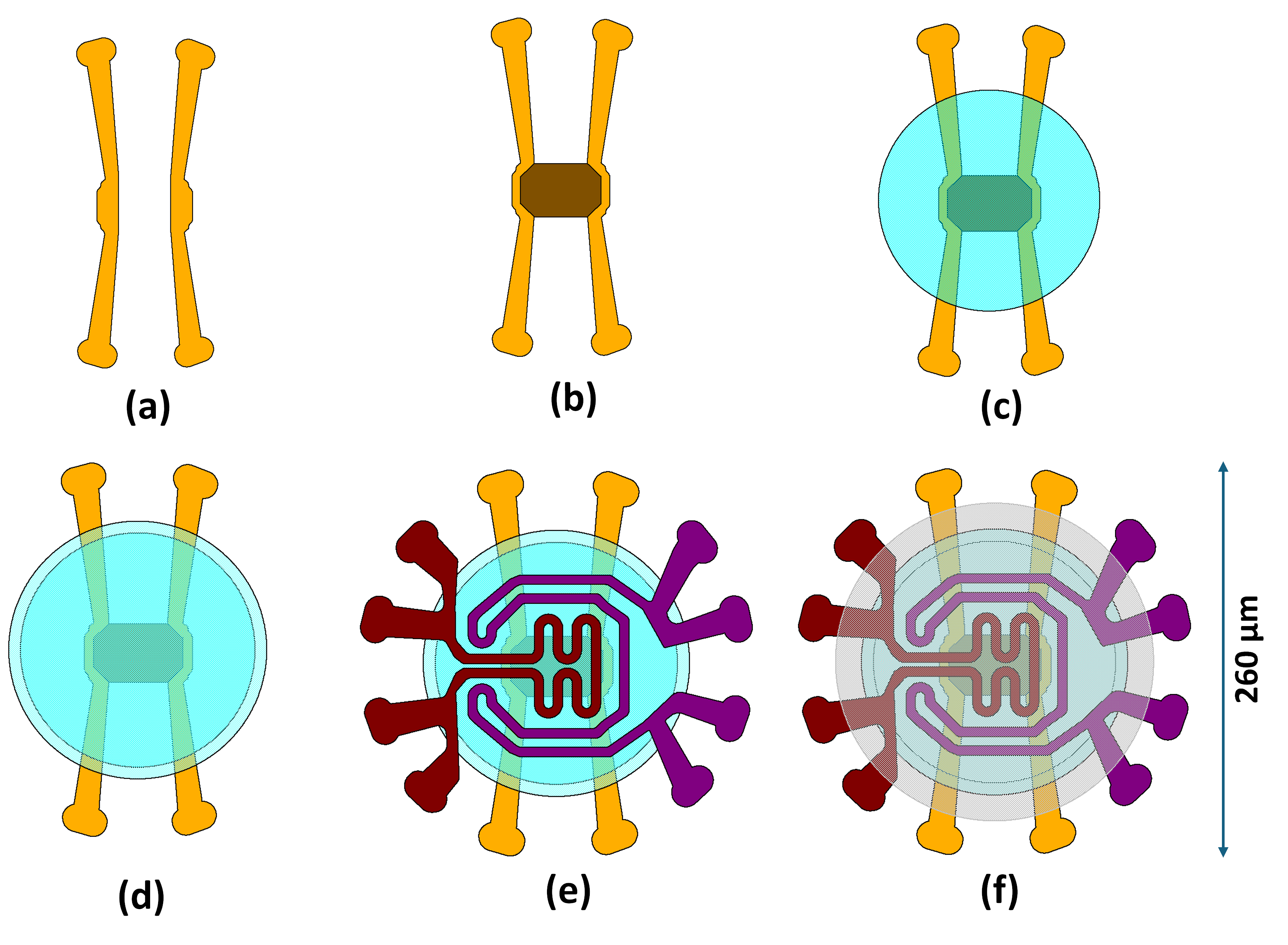}
  \caption{Schematic of the calorimeter layers in the order of deposition. (a) Thermometer leads (b) Thermometer (c) First insulation layer (d) Second insulation layer (e) AC (dark brown) and offset (dark purple) heaters (f) Final insulation layer.
  }
\label{fig:fabrication steps}
\end{figure}

The nanocalorimeters are built on a 4-inch silicon wafer with a thickness of 525$\pm$20 $\mu$m, coated with a 1~$\mu$m thick SiO$_2$ layer on both sides, by a batch fabrication process. Over 1600 devices can be fabricated on a single wafer. After depositing all layers, the calorimeters are divided into small chips through front and backside etching. The calorimeter consists of layers of thin-film thermometer, AC heater, and offset heater with a double layer of electrical insulation between the active layers. Figure~\ref{fig:fabrication steps} shows the schematic of the calorimeter layers in the order of deposition.
The AC heater is a meander-shaped resistor that covers the central area of the calorimeter to ensure uniform heat flow to the sample. It applies an oscillating current that induces temperature oscillations in the sample. The temperature modulated region has an area of approximately 50~$\mu$m diameter, enabling specific heat measurements of samples with corresponding size.
The offset heater lies around the central area. It is supplied with a DC current, which locally raises the temperature of the sample.
The double insulation layer avoids pinholes between layers and eliminates step-coverage issues.
The low thermal mass of the calorimeter stack ensures a comparatively small contribution to the addenda heat capacity even for a small sample that should ideally cover the thermometer and AC heater, particularly at low temperatures.

\subsection{Frontside fabrication}

Fabrication of each layer is done by photolithography, thin-film deposition by sputtering and e-beam evaporation, oxygen ashing, and double-layer lift-off. The thermometer is a 40~$\mu$m $\times $ 40 $\mu$m square resistor of thickness 65~nm. It is deposited onto a pair of leads composed of 20 nm Cr and 40 nm Au deposited using e-beam evaporation. The thermometer material is a mixture of a metal with short electronic mean free path and an insulating oxide, produced by co-sputtering NiCr and SiO$_2$ in an argon and oxygen atmosphere \cite{fortune2023wide}. Following the deposition, a 25 nm thick SiO$_2$ film is sputtered without breaking the vacuum to prevent the thermometer material from reacting further with the atmosphere.
This thin-film thermometer, in direct contact with the sample, provides high sensitivity across a wide temperature range and offers a rapid thermal response. The room temperature resistance of the thermometer can be tuned between 1200 - 2200~$\Omega$ to achieve the desired properties. 

Both heaters are made of 80~nm thick Cr, deposited in a single fabrication step. Careful mask alignment with the thermometer layer is needed to avoid overlap with sharp steps of the combined layer of thermometer and thermometer leads.

SiO$_2$ is used for the double-insulation layer, which is sputtered in two steps, with each layer having a thickness of 55~nm, as shown in Fig.~\ref{fig:fabrication steps}c and d.
The layers are covered by a thick SiO$_2$ layer from the top as a final step (Fig.~\ref{fig:fabrication steps}f). The thickness of each layer is chosen to reach full step-coverage of previous layers and to ensure that all edges have sufficiently thick insulation barriers to subsequent layers.

\subsection{Etching}

\begin{figure}
  \centering
\includegraphics[width=0.45\linewidth]{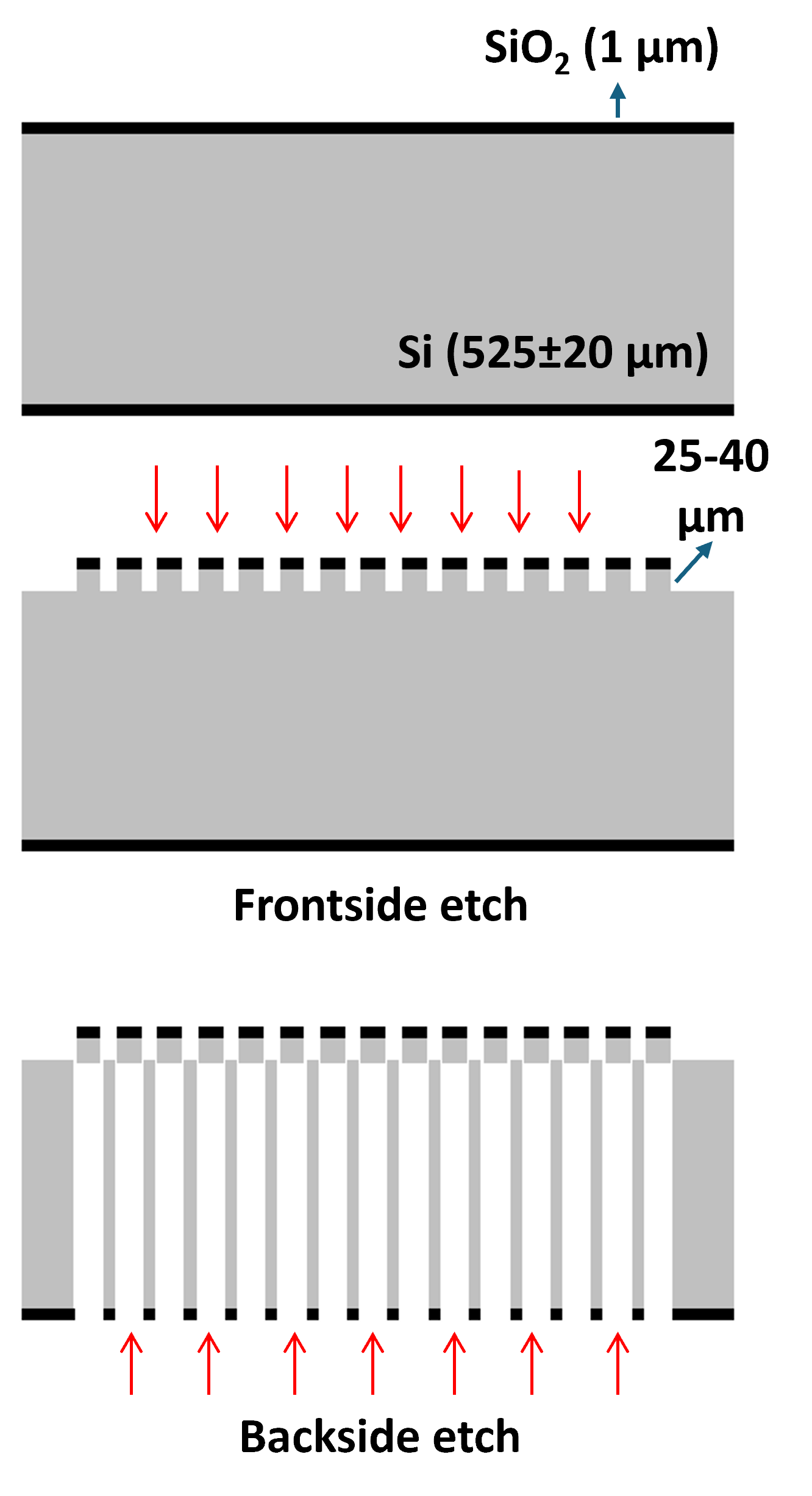}
  \caption{Schematic illustrating the etch steps, showing the full wafer, frontside etch, followed by backside etch. Red arrows indicate the direction of the plasma.
  }
\label{fig:etch diagram}
\end{figure}

The devices are separated into individual calorimeter chips through a dry etch process after frontside fabrication. The 525~$\mu$m thick silicon wafer undergoes anisotropic Bosch etch from both front and backside, resulting in thin calorimeter chips with diameters of 300~$\mu$m and thicknesses ranging from 25 to 40~$\mu$m. The schematic of the etch process is illustrated in Fig.~\ref{fig:etch diagram}.

The 4-inch wafer is initially etched from the frontside, followed by backside etch. For the frontside etch, a photoresist etch mask is created to cover the devices, leaving a circular ring around them. The SiO$_2$ on top is anisotropically etched at a rate of 400 nm/min. This is followed by Bosch etch of silicon to a depth of up to 40~$\mu$m, with a rate of 0.9~$\mu$m/cycle, each cycle taking 7 s. The etched depth is measured under a microscope by adjusting the focus. Figure~\ref{fig:etching results}a shows the device after frontside etch with the focus on the top surface.

The backside etch mask is a 100 nm thick Cr film created through e-beam evaporation with subsequent lift-off in acetone, prepared before the frontside etch. Chromium is selected for its lower etch rate compared to silicon. The backside mask covers the entire wafer except for discs of 800~$\mu$m diameter centered around the devices. For the backside etch, the frontside of the wafer is attached to a standard dicing tape (Blue Tack Roll from Semiconductor Equipment Corp.), to hold the final calorimeter chips. The process is otherwise similar to the frontside etch. After etching the SiO$_2$ covering, the silicon is etched until the ring around the device becomes visible, as shown in Fig.~\ref{fig:etching results}b. The process is halted once all the rings are clearly visible. A small variation in thickness could be expected due to variations in etch rate depending on the location on the wafer. After the etch process, the calorimeter chips, shown in Fig.~\ref{fig:etching results}c, can be collected and cleaned.

\subsection{Electrical connections}
\begin{figure}
  \centering
\includegraphics[width=1\linewidth]{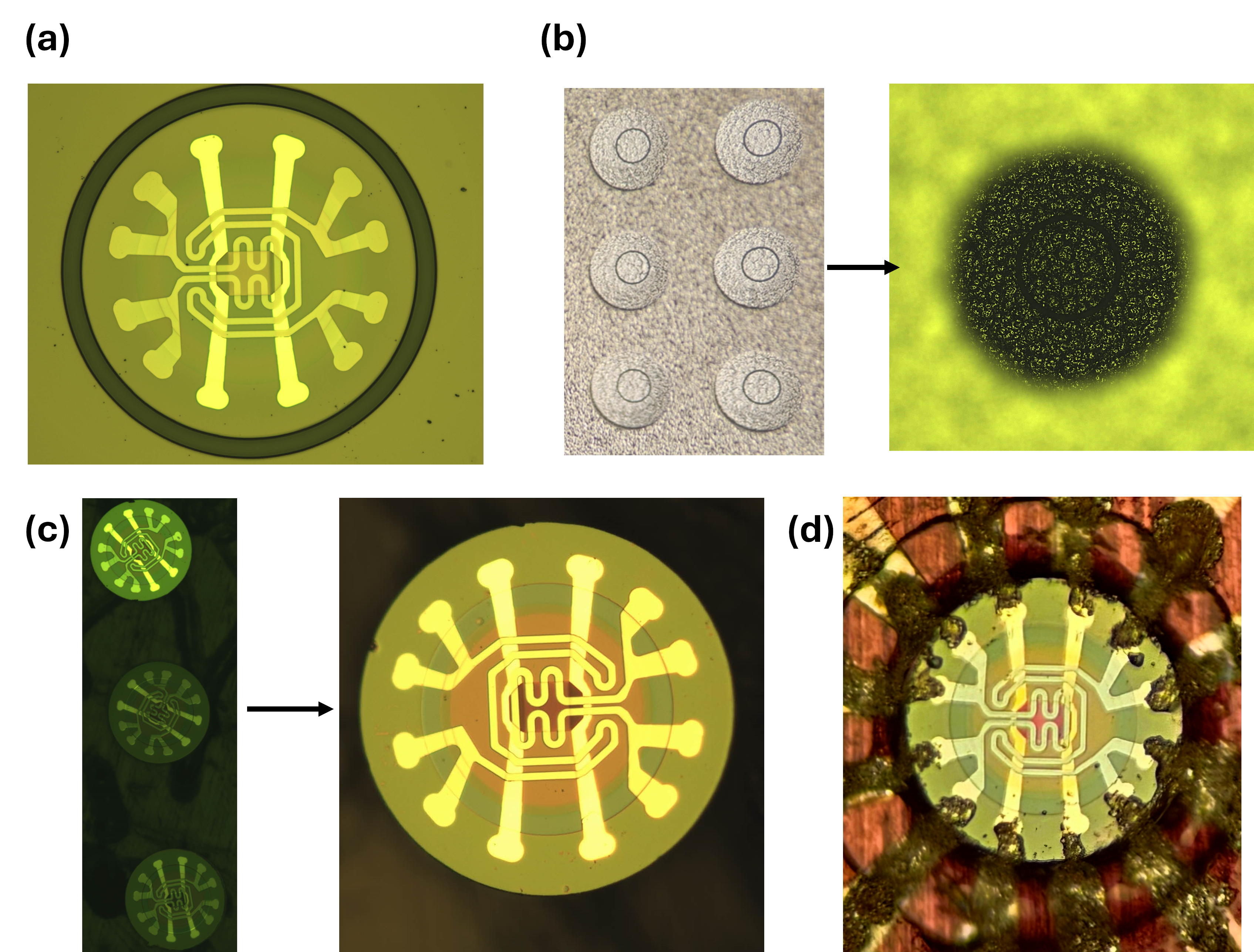}
  \caption{Results after etching from both sides. (a) Frontside mask featuring a ring around the device, which is etched through. (b) Backside of the wafer (zoomed view) after backside etch. The rings around the devices are visible which means the devices are separated. (c) Calorimeter chips after cleaning and separation. Magnified view of one chip is shown on the right side. (d) Calorimeter contact pads connected to external electrical leads using silver epoxy.
  }
\label{fig:etching results}
\end{figure}

The resistance of each active element is measured using the four-probe method, which eliminates contact resistance and contributions from the leads. 
The contact pads of the active layers, 12 in total, each have a width of 30 µm and can be connected to external electrical contacts using a conductive epoxy, either by hand or with a micro-manipulator, as depicted in Fig.~\ref{fig:etching results}d. The sidewall of the chip is insulated by melted crystal bond before making electrical contacts. The calorimeter chip is most easily integrated into a high-pressure cell using a split-gasket approach \cite{KondedanGasket}, but could also be connected to leads that are directly deposited onto the diamonds \cite{grzybowski1984band, gao2005accurate, weir2000epitaxial} or by other means.

\section{Thermometer characteristics}

The NiCrSiO$_x$ thermometers used here are based on the thin-film thermometers \cite{fortune2023wide} designed to replace the AuGe thermometers used previously in our membrane-based nanocalorimeters \cite{tagliati2012differential}.
A reduction of the thermometer size from the 80~$\mu$m used earlier is needed to fit the active layers into the miniaturized calorimeter. The optimal size of the thermometer depends on several factors. A large thermometer will require a correspondingly large heater to maintain temperature uniformity. A large heater requires a large sample to maintain the sample to addenda ratio and increases the heater power, which may be problematic at cryogenic temperatures. However, if the thermometer is very small, self-heating would become a significant problem at the lowest temperatures. The thermometer properties may also be affected by the thermometer leads and film uniformity.

\begin{figure}
  \centering
\includegraphics[width=1\linewidth]{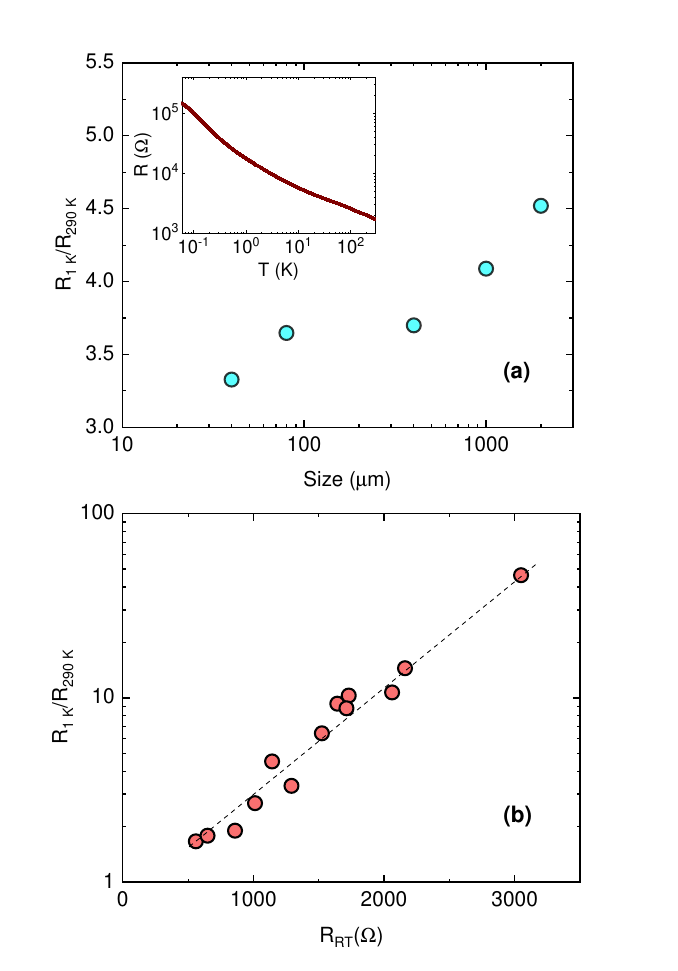}
  \caption{ (a) Variation of resistance ratio between 1~K and 290~K with respect to the thermometer size, for thermometers fabricated in the same deposition run. The inset shows the temperature-dependent resistance of a thermometer having a room-temperature resistivity of 10$^{-4}$ $\Omega$m.
  (b) The resistance ratio between 1~K and 290~K
  for different 40~$\mu$m thermometers as a function of corresponding room temperature resistance.
  }
\label{fig:diff size thermo}
\end{figure}

To investigate the impact of size on thermometer characteristics, thermometers with a size ranging from 40 to 2000~$\mu$m were fabricated under identical conditions. These thermometers exhibit square resistance at room temperature that vary within a range of 100~$\Omega$, and they show a resistance value that increases with decreasing temperature, as shown in the inset of Fig.~\ref{fig:diff size thermo}a. These thermometers have been calibrated following the method described in \cite{fortune2023wide}. 

Figure~\ref{fig:diff size thermo}a illustrates the relationship between thermometer size and the ratio of thermometer resistance from 1~K to 290~K. The ratio decreases slightly as the size is reduced, but the sensitivity remains largely unaffected.
The slight variation in the ratio between different sizes of thermometers could be attributed to the re-sputtering of gold from the leads into the thermometer region, particularly along the edges during thermometer deposition. This process results in a tendency to become slightly more metallic. This effect is more pronounced in smaller thermometers, resulting in a lower resistance ratio for given deposition conditions.

Figure~\ref{fig:diff size thermo}b shows the low-temperature sensitivity of several 40~$\mu$m thermometers as a function of room temperature resistance, with the variation of properties obtained by tuning the metal to insulator deposition conditions. The resistance ratio shows a nearly exponential increase with room temperature resistance, allowing for predictable low-temperature characteristics of the 40~$\mu$m thermometer, similar to those observed in the 80~$\mu$m thermometers \cite{fortune2023wide}. This confirms that miniaturizing the thermometer dimensions does not significantly alter the functional properties of the thermometers. 

\section{Heat-flow analysis}

\begin{figure}
  \centering
\includegraphics[width=0.9\linewidth]{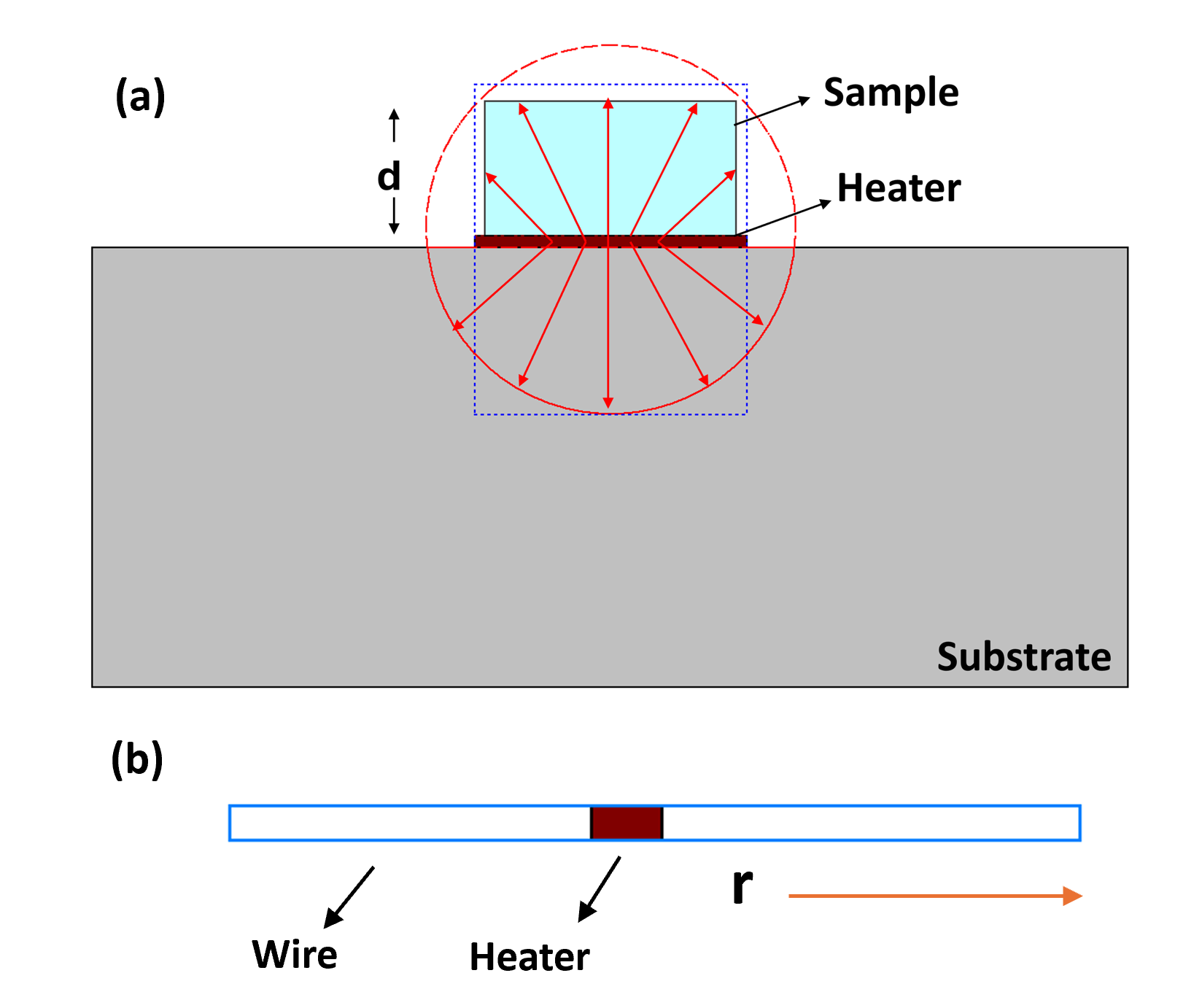}
  \caption{Schematic of the calorimeter with a sample on top. The red arrows depict the direction of heat flow from the heater, while the red circle indicates the extent of diffusion ($l_\mathrm{{th}}$). The dotted blue box highlights the volume that is temperature modulated in the 1D diffusion model. (b) Illustration of a thin wire with a heat source located at a single point, where oscillations propagate along the \textit{r} direction.
  }
\label{fig:heat flow diagram}
\end{figure}

The heat flow of the calorimeter system can be modeled as a one-dimensional heat diffusion problem by approximating the planar AC heater geometry illustrated in Fig.~\ref{fig:heat flow diagram}a as a 1D wire with a point heat source of Fig.~\ref{fig:heat flow diagram}b. In a wire, the heat wave propagates along the wire (in the \textit{r} direction) from the point heat source. In the calorimeter system, although heat is transmitted in multiple directions, as indicated by the red arrows in the Fig.~\ref{fig:heat flow diagram}a, the primary region of interest for heat propagation
from the heater area is up and down, enclosed by the dotted blue box in the figure, which includes the sample and a portion of the silicon substrate. 

The AC heater generates a modulated heat wave, which transfers heat to both the sample and the silicon substrate. The extent of the thermal wave is determined by the angular frequency $\omega$ of the temperature oscillation.
In the 1D model, temperature modulation as a function of distance (\textit{r}) is given by \cite{hatta1985thermal}
\begin{equation} \label{T(r)}
T_\mathrm{{ac}}(r) \propto \frac{1}{\omega}\exp(-kr-ikr-i\pi/2),
\end{equation}
where the decay constant \textit{k} is the inverse of the thermal diffusion length $l_\mathrm{{th}}$. Equation~(\ref{T(r)}) is valid under the condition that $\omega\tau_\mathrm{e} \gg 1$, where $\tau_\mathrm{e}$ is the relaxation time between the system and the thermal bath. 

The thermal diffusion length $l_\mathrm{{th}}$ can be estimated within the pseudo-one-dimensional geometry, and is given by
\begin{equation} \label{eq:thermal length}
l_\mathrm{{th}}(\omega) = \sqrt{\frac{2D}{\omega}}.
\end{equation}
Here, \textit{D} is the diffusivity of the material, defined as
\begin{equation} \label{diffusivity}
D = \frac{K}{\varrho c_\mathrm{p}},
\end{equation}
where \textit{K}, $\varrho$, and \textit{c}$_\mathrm{p}$ represent thermal conductivity, volume density, and specific heat respectively. For uniform heating within the sample, the sample thickness \textit{d} must be much smaller than $l_\mathrm{{th}}$. Otherwise, a thermal gradient will develop in the sample perpendicular to the planar heat source.
Beyond the range of the thermal diffusion length (\mbox{\textit{r} $\gg$ $l_\mathrm{{th}}$}), the temperature oscillation is negligible and the substrate (including possible pressure medium) acts as the thermal bath.

The diffusion introduces a frequency-dependent heat capacity, $C = C(\omega)$ that is probed. To accurately measure the sample heat capacity, additional addenda heat capacity must be minimized and the full sample heat capacity must be included. This optimization is non-trivial and depends on the properties of the sample, substrate, and pressure medium. By adjusting the thermal diffusion length through the frequency, using Eq.~(\ref{eq:thermal length}), the thermal oscillation should be restricted to a volume that extends only slightly beyond the sample. The relative contributions from sample and substrate to the probed heat capacity then depends on their effusivities $e = (K \varrho c_\mathrm{p})^{1/2}$, since the heater area $A_\mathrm{heat}$ defines a probed volume $l_\mathrm{th}A_\mathrm{heat}$ on each side of the heater so that the participating heat capacity is given by $l_\mathrm{th}A_\mathrm{heat}\varrho c_\mathrm{p} \propto e$. Thus, when the sample effusivity exceeds that of the substrate, a relatively small substrate addenda will be probed, leaving the sample heat capacity well defined. Similarly to the effusivity of the calorimeter substrate, the effusivity of the pressure medium should be kept as low as possible, to restrict the thermal wave from extending beyond the top surface of the sample. This is normally the case, since most pressure media are insulators with low effusivities. In this case, the thermal oscillation is localized well to the sample and some part of the substrate. Since the thermal wave never reaches the metallic pressure cell gasket, the probed volume remains small and well defined, representing the sample heat capacity with a frequency dependent background from the silicon substrate.

\section{Calorimeter performance}
\subsection{Frequency and temperature dependence}
\begin{figure}[t!]
  \centering
\includegraphics[width=1\linewidth]{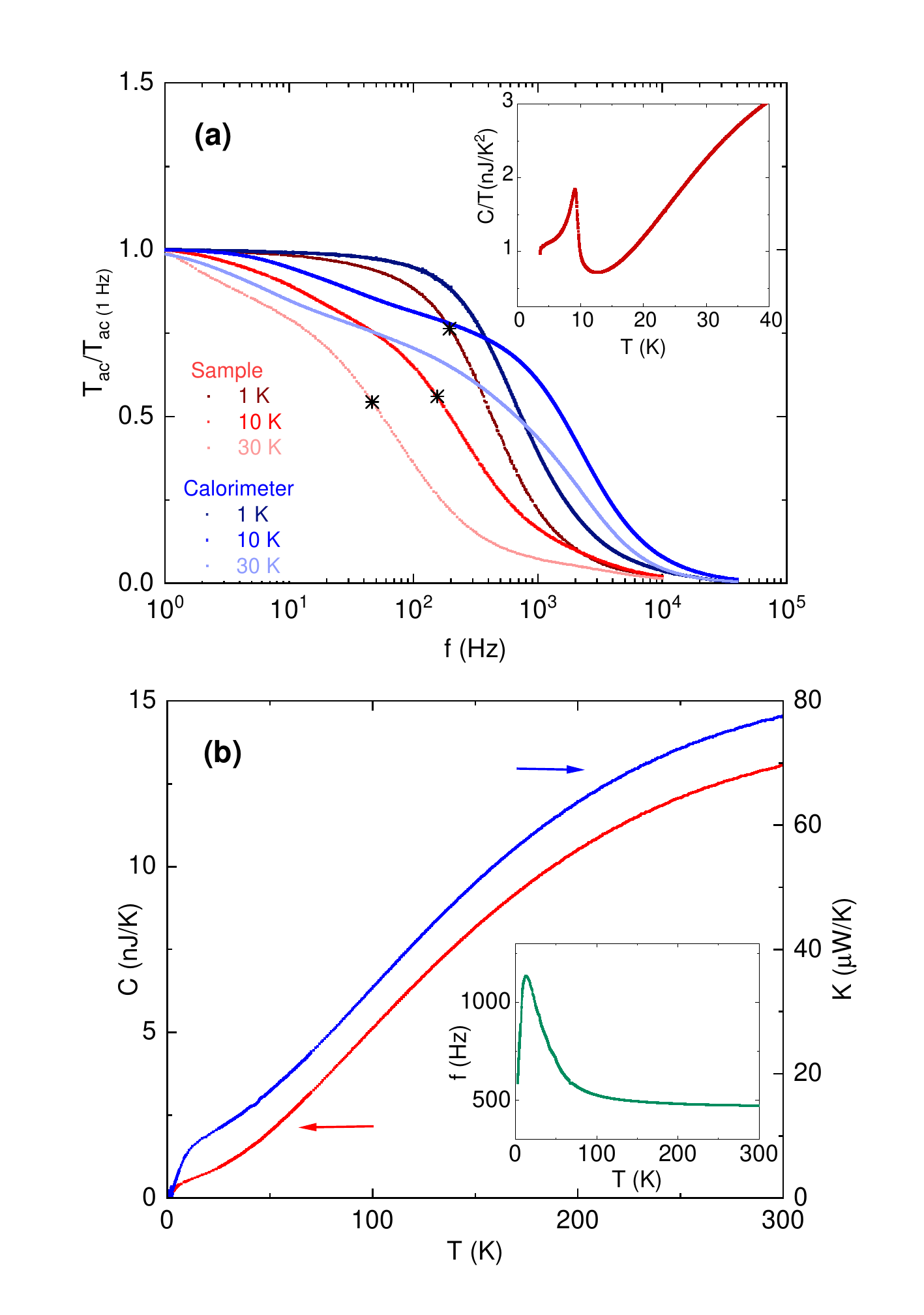}
\caption{Calorimeter characterization. (a) Frequency dependence of the temperature oscillation amplitude, normalized to oscillation amplitude at 1 Hz, for calorimeter with and without sample, measured at 1~K, 10~K, and 30~K. The inset shows the measured heat capacity, plotted as $C/T$, as a function of temperature for a Gd$_2$Zn$_{17}$ crystal. The measurement frequency was varied to maintain a constant phase between applied power and temperature oscillation. The corresponding measurement frequencies at 1~K, 10~K, and 30~K are indicated by black markers in the main figure. (b) Thermal conductance and heat capacity of the calorimeter chip, from 0.1~K to 300~K. The inset shows the corresponding frequency during the measurement. }
\label{fig:calorimeter characterization}
\end{figure}

The empty calorimeter was characterized from room temperature to below 1~K. A 0.77\,$\mu$g Gd$_2$Zn$_{17}$ crystal with an antiferromagnetic transition at 9 K \cite{marquina1996specific} was also used as a test sample, mounted on the calorimeter using a micromanipulator with a tiny droplet of Apiezon-N grease for thermal attachment. Figure~\ref{fig:calorimeter characterization}a shows the temperature oscillation amplitude of the empty calorimeter (red curves) and calorimeter with sample (blue curves) at different temperatures, as a function of frequency. From the figure, one can see that the temperature oscillation is significantly reduced by the presence of the sample, indicating that the sample heat capacity acts to reduce the thermal oscillation.

Analyzing the frequency dependence of Fig.~\ref{fig:calorimeter characterization}a, three different frequency regimes can be identified for each curve. At low frequency, the temperature oscillation amplitude is constant (at 1 K) or slightly decreasing (at 10 K and 30 K) with frequency. This behavior corresponds to the low-frequency regime of regular ac calorimetry, where the effective thermal link $K_\mathrm{eff}$ is probed and there is no effect from the sample heat capacity on the temperature oscillation; $T_\mathrm{ac} \approx P_\mathrm{ac}/K_\mathrm{eff}$. In this regime the phase between temperature oscillation and applied power $P_\mathrm{ac}$ is small.

At intermediate frequencies, the temperature oscillation starts to drop quickly  with frequency, indicating that the heat capacity reduces the temperature oscillation. For a well-defined heat capacity $C_\mathrm{eff}$, the relation between $C_\mathrm{eff}$ and $T_\mathrm{ac}$ is given by \cite{tagliati2012differential}
\begin{equation}\label{TacofC}
C_\mathrm{eff}=\frac{P_\mathrm{ac}}{\omega T_\mathrm{ac}}\sin \varphi ,
\end{equation}
where $\varphi$ is the phase between applied power and temperature oscillation. Since the calorimeter contribution to the heat capacity will be frequency dependent, it is not trivial to acquire absolute accurate measurements of the sample heat capacity $C_\mathrm{s}=C_\mathrm{eff}(\omega)-C_\mathrm{cal}(\omega)$.

Going to the highest frequency range, the effective sample heat capacity will start to decrease due to the thermal length becoming shorter than the sample thickness. In this range, $\tan \varphi = \omega C_\mathrm{eff}/K_\mathrm{eff}$ is no longer proportional to $\omega$, but starts to quickly decrease with increasing frequency with a concurrent increase of $K_\mathrm{eff}(\omega)$ and decrease of $C_\mathrm{eff}(\omega)$. There is, thus, a maximum in $\tan \varphi$ beyond which absolute accuracy cannot be obtained. The maximum in $\tan \varphi$ depends on the sample heat capacity, sample thermal conductivity, and how well the sample is attached to the calorimeter, but is normally located near $\tan \varphi \approx 1$.

The suitable measurement frequency depends on the requirements on absolute accuracy and resolution. Measurements in the highest frequency range, beyond the maximum in $\tan \varphi$ will give higher resolution (other parameters optimized). As seen in Fig.~\ref{fig:calorimeter characterization}a the ratio of $T_\mathrm{ac}$ without and with sample at 30 K reaches a factor of 6 at 1\,kHz, indicating that the probed $C_\mathrm{s}(\omega)$ is about 5 times higher than $C_\mathrm{cal}(\omega)$ at this frequency. However, measurements at the intermediate frequency range will always include the full $C_\mathrm{s}$ (temperature oscillation is uniform in the entire sample), retaining the possibility to obtain absolute accuracy as long as the calorimeter $C_\mathrm{cal}(\omega)$ is separately measured.

The inset of Fig.~\ref{fig:calorimeter characterization}a shows the heat capacity of the small Gd$_2$Zn$_{17}$ crystal. The antiferromagnetic ordering transition at 9 K is clearly observed. These measurements were taken with $\tan \varphi = 0.8$. The corresponding locations of the measurement frequencies in the main panel of Fig.~\ref{fig:calorimeter characterization}a are indicated by black markers.

Figure~\ref{fig:calorimeter characterization}b shows the temperature dependence of the calorimeter heat capacity $C_\mathrm{cal}$ and thermal conductance $K_\mathrm{eff}$ of the empty calorimeter chip. The temperature dependence of the measurement frequency is shown in the inset. This frequency was obtained by continuously adjusting frequency to obtain $\tan \phi = 1$. The $C_\mathrm{cal}$ curve gives an indication of the smallest samples that can be studied with the calorimeter. The background corresponds roughly to a 30 ng piece of copper, making it possible to obtain signals from samples down to the few ng range.

\subsection{Pressurization in a diamond anvil cell}

The calorimeter chips were subjected to high-pressure conditions using a diamond anvil cell with a culet diameter of 1.2 mm. During these tests, the devices were fully immersed in the pressure medium to ensure uniform pressure distribution. In our pressure cell, the devices withstood pressures up to the maximum cell pressure of 2 GPa without failure. This ability to withstand pressure confirms that the chosen calorimeter substrate material survives in high-pressure experiments. The ultimate maximum pressure tolerable by the silicon chips remains to be determined.

\section{Summary and conclusions}
In summary, we describe a miniaturized chip calorimeter for heat capacity measurements at low temperatures and high pressures. The miniaturization enables the calorimeter to fit into small sample volumes, such as inside high pressure diamond anvil cells combined with cryogenic systems. From an analysis of the frequency dependence of the calorimeter with and without a test sample, we find the suitable measurement conditions, where the sample heat capacity can be separated from the frequency dependent calorimeter background. While the calorimeter is designed for high pressure studies, it could also find its applications for high-frequency measurements of small samples, such as in pulsed magnetic fields. 

\section*{Acknowledgements}
Support from the Knut and Alice Wallenberg Foundation
under grant number KAW 2018.0019 and
the Swedish Research Council, grant number 2021-04360, are acknowledged. We thank J.\,Palmer-Fortune, A.\,Bangura, U.\,H{\"{a}}ussermann, and A.\,Khansili for insightful discussions on the development of the calorimeter.

\appendix



\bibliographystyle{elsarticle-num} 
\bibliography{Kondedan}






\end{document}